\documentclass[11pt, oneside]{article}   	% use "amsart" instead of "article" for AMSLaTeX format
\usepackage{geometry}                		% See geometry.pdf to learn the layout options. There are lots.
\usepackage{amssymb}
\usepackage{mathtools}
\usepackage{color}

\newcommand {\bp}{{\bf p}}
\newcommand {\bq}{{\bf q}}
\newcommand {\uu}{\underline}
\newcommand {\bra}{\langle}
\newcommand {\ket}{\rangle}

\author		{I. Frolov\\
Department of Mathematics,\\
Moscow Engineering Physics Institute\\
115409, Kashirskoe shosse 31, Moscow,\\
frolovi55@mail.ru\\
						A. Schwarz\\
Department of Mathematics\\
University of California\\
Davis, CA 95616, USA,\\
schwarz @math.ucdavis.edu\\}

\title{Functional integrals in geometric approach to quantum theory}
\date{}							% Activate to display a given date or no date

\begin{document}
\maketitle

\begin {abstract}

In quantum mechanics, one can express the evolution operator and other quantities in terms of functional integrals. The main goal of this paper is to prove corresponding results in geometric approach to quantum theory. We apply these results to the formalism of L-functionals.
\end {abstract}
{\bf Keywords}  functional integral; Banach space; geometric approach

MISC 81T

\section {Introduction}
In geometric approach \cite {GA1},\cite {GA},\cite {FRS}  the evolution operator of physical system obeys the equation of motion
\begin {equation}
\label {EM}
\frac {d\sigma}{dt}=H(t)\sigma (t),
\end {equation}
where $H(t)$ is a linear operator acting in Banach space (or, more, generally,  in complete topological vector space) $\cal L$. We say that $H(t)$ is the "Hamiltonian" of the physical system. In what follows we assume that
$H(t)=H$ does not depend on time $t$. This condition is imposed only to simplify notations; all results can be proved also for time- dependent "Hamiltonian".

In the standard approach to quantum mechanics the evolution operator acts in Hilbert space; it obeys the equation
(\ref {EM}) where $H(t)$ is a skew-adjoint operator. It can be represented by a functional integral. One of the ways to obtain such a representation is based on the notion of a symbol of an operator; this way was suggested by F. Berezin \cite {BER} (see \cite {BS} for details). We use the same ideas to obtain a representation of the evolution operator in  Banach  spaces (or in topological vector spaces)   in terms of functional integrals.  Such a representation was considered in numerous mathematical papers. (See, in particular,\cite {ZA},  \cite {SMI}, \cite {NEK}, \cite {NEK1}. Our approach is close to the ideas of these papers.) Our main result is a construction of functional integrals in the formalism of L-functionals (Section  5).

A symbol of an operator $A$ is a function $\underline A$ defined on some measure space. It should depend linearly of $A$. We assume that the symbol   of the identity operator $1$ is equal to $1$ and the composition of operators corresponds to the operation on symbols denoted by $*:$ if $C=AB$ then $\underline C=\underline A *\underline B.$

The simplest way to construct symbols of operators in quantum mechanics is  to use the fact that the Fourier transform of delta-function is a constant. The matrix (the kernel in the language of 
mathematics) of the unit operator is $\langle \bq_2|1|\bq_1\rangle=\delta (\bq_1-\bq_2)$ in coordinate representation and $\langle \bp_2|1|\bp_1\rangle=\delta(\bp_1-\bp_2)$ in momentum representation. Taking  Fourier transform of  matrix $\langle \bq_2|A|\bq_1\rangle$ of the operator $A$ with respect to the variable $\bq_1-\bq_2$
 we obtain 
$p-q$ symbol:

$$\underline A^{p-q} (\bp,\bq)= \int d{\bf y}\langle {\bf y}|A|\bq\rangle e^{i\bp(\bq-{\bf y})}.$$

Similarly taking Fourier transform of $\langle \bp_2|A|\bp_1\rangle$ with respect to variable $\bp_1-\bp_2$ we obtain $q-p$-symbol:
{
$$\underline A^{q-p} (\bq,\bp)= \int d{\bf y}\langle {\bf y}|A|\bp\rangle e^{-i\bq(\bp -{\bf y})}.$$

If  $A$ is a differential operator with polynomial coefficients we can express it as a polynomial of operators  $\hat q^j$  (operators corresponding to the coordinates $q^j$)  and $\hat p_j=\frac {1}{i}\frac {\partial}{\partial q^j}$  (momentum operators)  Representing $A$ in $q-p$ form (coordinate operators from the left of momentum operators) and "removing hats" we obtain $q-p$-symbol.
Representing $A$ in $p-q$-form (placing momentum operators from the left of  coordinate operators)  and "removing hats" we get $p-q$ symbol.

Notice that in our notations $\hbar=1.$ Sometimes it it is convenient to consider  families of symbols $ \underline A^{q-p}_{\hbar } (\bq,\bp)$ and $ \underline A^{p-q}_{\hbar } (\bp,\bq)$ depending on parameter $\hbar.$ 

We will describe a very general construction of symbols. We will use this construction to represent physical quantities in terms of 
functional integrals.  Our results generalize the results by F. Berezin \cite {BER}, \cite {BS}  proved for operators in Hilbert spaces.
They can be applied also to coherent states \cite {PER} and their generalizations.

We did not try to give rigorous proofs of our results, however imposing some conditions one can  make our exposition rigorous (for example, using the ideas of \cite {ZA}, \cite {SMI}, \cite {NEK}).
We use our general results to obtain a representation of physical quantities in terms of functional integrals   in the formalism of $L$-functionals \cite {SCH}, \cite {S}, \cite {MO}.

\section {Functional integrals} 

Let us consider  symbols of  linear operators acting in the space $\cal L.$

The evolution operator can be represented in the form 
$$ \sigma (t)= e^{tH}=\lim _{N\to \infty} (1+\frac {tH}{N})^N.$$
For $N\to \infty$ the symbol of the  operator $1+{\frac {tH}{N}}$ can be approximated  by $\exp{\frac {t}{N}}\underline H$:
$$\underline{1+\frac {tH}{N}}=e^{\frac {t}{N}\underline H}+O(N^{-2}).$$ Using this relation we obtain an expression for the symbol of the evolution operator;
\begin {equation}\label {EE}
\underline{\sigma (t)}= \lim_{N\to \infty}I_N(t),
\end {equation}
where
$$I_N(t)=e^{\frac {t}{N}\underline H}*...*e^{\frac {t}{N}\underline H}$$
($N$ factors).

In many cases $I_N(t)$ can be interpreted as an approximation to a functional integral. Notice, however, that even without this interpretation we can apply the Laplace or stationary phase method to the calculation of $I_N(t)$. This allows us to obtain some results that often are obtained in the language of functional integrals without using this language.

Let us consider a class of symbols generalizing  $q-p$ symbols and Wick symbols of quantum mechanics.

We start with complete topological vector space $\cal L.$

We assume that the symbol of an operator  $A$ acting in $\cal  L$ is a  (generalized)  function $\underline A(\alpha,\beta')$ of two variables  (a function on ${\cal M}\times \cal M'$) and that the
 symbol of the product $C=AB$ of operators $A$ and $B$ can be expressed in terms of the symbols of operators $A$ and $B$ by the formula
\begin{equation}\label {NN}
\underline C(\alpha,\beta')=\int d\gamma d\gamma'\underline  B(\alpha,\gamma')R(\gamma,\gamma')\underline A(\gamma,\beta')e^{c(\alpha,\gamma')+c(\gamma,\beta')-c(\alpha,\beta')},
\end {equation}
where $c(\gamma,\gamma'),R(\gamma,\gamma')$ are functions  on $\cal M\times \cal M'$ and $d\gamma d\gamma'$ is the measure on this space.  (We assume that  $\cal M\times \cal M'$ is a measure space. This assumption can be weakened to allow infinite-dimensional spaces with integration defined for some class of functions on these spaces.)
Taking $A=B=C=1$ in (\ref {NN}) we obtain the following restriction on these functions:
\begin{equation}\label {c11}
1=\int d\gamma d\gamma' e^{c(\alpha,\gamma')+c(\gamma,\beta')-c(\alpha,\beta')}R(\gamma,\gamma').
\end {equation}

If $R(\gamma,\gamma')$ is  represented in the form 
$$R(\gamma,\gamma')=e^{-r(\gamma,\gamma')}$$
 the formulas (\ref{NN},\ref{c11}) can be rewritten as 

\begin{equation}\label {NN1}
\underline C(\alpha,\beta')=\int d\gamma d\gamma'\underline  B(\alpha,\gamma')\underline A(\gamma,\beta')e^{c(\alpha,\gamma')+c(\gamma,\beta')-c(\alpha,\beta')-r(\gamma,\gamma')},
\end {equation}
\begin{equation}\label {c12}
1=\int d\gamma d\gamma' e^{c(\alpha,\gamma')+c(\gamma,\beta')-c(\alpha,\beta')-r(\gamma,\gamma')}.
\end {equation}

It follows that the symbol $\underline {C}(\alpha,\gamma')$ of the product $C$ of $N$  operators $A_1,...,A_N$ is given by the formula

{
\begin {equation}\label{N}
\underline {C}(\gamma,\gamma')= \int d\gamma_1d\gamma'_1...d\gamma_{N-1}d\gamma'_{N-1}\underline A_N(\gamma,\gamma'_{N-1})...\underline A_2(\gamma_2,\gamma'_1)\underline A_1(\gamma_{1},\gamma') e^{\rho_N},
\end {equation}
}
where
\begin{equation}
\rho_N=c(\gamma,\gamma'_{N-1})+c(\gamma_{N-1},\gamma'_{N-2})+...+ c(\gamma_{1},\gamma') -c(\gamma,\gamma')-r(\gamma_1,\gamma'_1)-...-r(\gamma_{N-1},\gamma'_{N-1}).
\end{equation}

We see that in our case
\begin {equation}\label {I}
I_N(t)=\int d\gamma_1d\gamma'_1...d\gamma_{N-1}d\gamma'_{N-1}e^{\frac {t}{N}(\underline H(\gamma,\gamma'_{N-1}) +\underline H(\gamma_{N-1},\gamma'_{N-2})+...+\underline H(\gamma_{1},\gamma'))}e^{\rho_N}. 
\end{equation}

Notice that assuming that operators at hand have trace  we can express the trace in terms of symbols:
\begin{equation} \label{TR}
  Tr A=\int d\alpha d\beta' \uu A(\alpha,\beta') e^{\tau (\alpha,\beta')},
\end{equation}
where $\tau (\alpha,\beta')=c(\alpha,\beta')-r(\alpha,\beta')$.
To justify this formula we verify that $ Tr AB=Tr BA$  using (\ref {NN1}). (A  trace on an algebra is defined as a linear functional that vanishes on commutators. We are checking that the RHS of (\ref {TR}) is a trace in this general sense. It seems that in  our situation this property specifies the trace up to a numerical factor.)

Using (\ref {TR})  we obtain that
  \begin{equation}\label {TRH}
  Tr e^{tH}= \lim_{N\to\infty}  J_N(t),
  \end {equation}
  where
\begin {equation}\label {TRJ}
J_N(t)=\int d\gamma d\gamma' d\gamma_1d\gamma'_1...d\gamma_{N-1}d\gamma_{N-1}\exp(\frac {t}{N}(\underline H(\gamma,\gamma'_{N-1}) +...+\underline H(\gamma_{1},\gamma')))e^{\tilde{\rho}_N}, 
\end{equation}
  
\begin{equation}\label {TRR}
\tilde{\rho}_N=c(\gamma,\gamma'_{N-1})+...+ c(\gamma_{1},\gamma') -r(\gamma,\gamma')-r(\gamma_1,\gamma'_1)-...-r(\gamma_{N-1},\gamma'_{N-1}).
\end{equation}

In the case when $c(\alpha,\beta')$ is a quadratic function  one can prove that one of 
solutions of the relation (\ref{c12})
has the form 
$$c(\alpha,\beta')=r(\alpha,\beta')+const,$$
where the constant can be absorbed in the definition of the measure $d\gamma d\gamma'$.        

Let us show that one can use (\ref {NN1}) to express the symbol of the evolution operator in terms of functional integrals.  We assume that $\cal M$ and $\cal M'$ are smooth manifolds, the function $c$ is differentiable, and $r=c$). Then the symbol $\underline{\sigma (t)} (\gamma,\gamma') $ of operator $ \sigma (t)= e^{tH}$ can be represented as a functional integral:  
\begin{equation} \label {SS}
I(\gamma, \gamma') =\int \prod d\gamma(\tau) d\gamma'(\tau)  e^{S[\gamma(\tau), \gamma'(\tau)]},
\end{equation}
where

\begin {equation} \label {SSS}
\begin{gathered}
 S[\gamma(\tau), \gamma'(\tau)]=S_0[\gamma(\tau), \gamma'(\tau)]{+ c(\gamma(t), \gamma') - c(\gamma, \gamma')},
\end{gathered}
 \end {equation}}
\begin {equation} \label {SSS_0}
\begin{gathered}
 S_0=\int_0^t \left(\uu H[\gamma(\tau), \gamma'(\tau)]-\dot \gamma'(\tau)\frac{\partial}{\partial \gamma'(\tau)}c(\gamma(\tau), \gamma'(\tau))\right)d\tau.    
\end{gathered}
 \end {equation}
 
This integral depends on $\cal M$-valued function $\gamma(\tau)$ and $\cal M'$-valued function $\gamma'(\tau)$. 
Here 
$0\leq \tau\leq t$ and we integrate over the set of  functions obeying boundary conditions $\gamma(0)=\gamma, \gamma'(t)=\gamma'.$ To prove this statement we notice that the expression
$$\frac {t}{N}(\underline H(\gamma,\gamma'_{N-1}) +\underline H(\gamma_{N-1},\gamma'_{N-2})+...+\underline H(\gamma_{1},\gamma'))+\rho_N $$
approximates the integral sum for the integral (\ref {SSS_0}).  (To define the functional integral we represent it as a limit of finite-dimensional integrals. This definition depends on the choice of approximation of functional integral by finite-dimensional integrals.)

The last two terms in (\ref {SSS}) cancel in the formula for the trace of the operator $e^{tH}$. Using this remark or formulas (\ref {TRH}-\ref {TRR}) we obtain   
\begin{equation} \label {SS2}
Tr(e^{tH})= \int \prod d\gamma(\tau) d\gamma'(\tau)  e^{S_0[\gamma(\tau), \gamma'(\tau)]},
\end{equation}
where $S_0$ is given by the formula (\ref{SSS_0}). 

(We integrate over the set of functions obeying  boundary conditions $\gamma(0)=\gamma(t), \gamma'(0)=\gamma'(t).$)

It is easy to check that the formula (\ref {NN}) is valid for $p-q$-symbols in $n$-dimensional space with:
{
$$\gamma=\bp;\;\;\gamma'=\bq;\;\; c(\gamma,\gamma')=-i(\bp,\bq);\;\;r(\gamma,\gamma')=-i(\bp,\bq);\;\;  d\gamma d\gamma'=d^n\bp d^n\bq /(2\pi)^n $$
}
and for $q-p$-symbols with:
$$\gamma=\bq;\;\;\gamma'=\bp;\;\; c(\gamma,\gamma')=i(\bp,\bq);\;\;r(\gamma,\gamma')=i(\bp,\bq). $$
 
 Using (\ref{I}) we can get functional integrals of quantum mechanics.
 
 One more case when it is possible to obtain functional integrals from (\ref {I}) and (\ref {TRH})  is the situation when $\cal M=\cal M'$, $R (\alpha,\beta')=\delta (\alpha, \beta')$ and $c(\gamma,\gamma)\equiv 0$. In this situation we have 
\begin {equation}\label {I2}
I_N(t)=\int d\gamma_1...d\gamma_{N-1}\exp(\frac {t}{N}(\underline H(\gamma,\gamma_{N-1}) +\underline H(\gamma_{N-1},\gamma_{N-2})+...+\underline H(\gamma_{1},\gamma')))e^{\rho_N}, 
\end{equation}
where
\begin{equation}
\rho_N=c(\gamma,\gamma_{N-1})+c(\gamma_{N-1},\gamma_{N-2})+...+ c(\gamma_{1},\gamma') -c(\gamma,\gamma')).
\end{equation}
When $N \to \infty$
$$c(\gamma,\gamma')= c(\gamma,\gamma+\Delta \gamma)\approx \Delta \gamma\frac{\partial}{\partial \tilde{\gamma}}c(\gamma,\tilde{\gamma})|_{\tilde{\gamma}=\gamma}$$ 
\begin {equation}
I_N(t)\simeq\int d\gamma_1...d\gamma_{N-1}\exp(\frac {t}{N}\sum_i(\underline H(\gamma_i,\gamma_i)+\frac {t}{N}\sum_i { \dot \gamma_i}\frac{\partial}{\partial \tilde{\gamma}}c(\gamma_i,\tilde{\gamma})|_{\tilde{\gamma}=\gamma_i}-c(\gamma,\gamma')).  
\end{equation}
It follows that the symbol $\underline{\sigma (t)} (\gamma,\gamma') $ of operator $ \sigma (t)= e^{tH}$ can be represented as a functional integral  
\begin{equation} \label {SS4}
I(\gamma, \gamma') =\int \prod d\gamma(\tau)   e^{S[\gamma(\tau)]},
\end{equation}
where 
\begin {equation} \label {SSS4}
 S[\gamma(\tau)]=\int_0^t \left(\uu H[\gamma(\tau), \gamma(\tau)]+\dot \gamma(\tau)\frac{\partial}{\partial \tilde{\gamma}}c(\gamma(\tau), \tilde{\gamma})|_{\tilde{\gamma}=\gamma(\tau)}\right)d\tau  { - c(\gamma, \gamma')}
 \end {equation}
with $\gamma(0)=\gamma, \gamma(t)=\gamma'.$ Similarly in this case
\begin{equation} \label {S444}
Tr(e^{tH})= \int \prod d\gamma(\tau)  e^{S_0[\gamma(\tau)]},
\end{equation}
where 
\begin {equation} 
 S_0[\gamma(\tau)]=\int_0^t \left(\uu H[\gamma(\tau), \gamma(\tau)]+\dot \gamma(\tau)\frac{\partial}{\partial \tilde{\gamma}}c(\gamma(\tau), 
 \tilde{\gamma})|_{\tilde{\gamma}=\gamma(\tau)}\right)d\tau. 
\end {equation}

\section { Covariant and contravariant  symbols}

Let us consider Banach spaces $\cal L$ and $\cal L'$  and systems of vectors
$e_{\alpha}\in \cal L$, $e'_{\beta'}\in \cal L'.$ Here  $\alpha\in \cal M, \beta'\in\cal M'$, $\cal M\times \cal M'$ is a measure space. (Again we can consider a more general case when   $\cal M\times \cal M'$ is a space with integration defined for some class of functions on this space.)  We assume that linear combinations  of vectors $e_{\alpha}$ are dense in $\cal L$ and linear combinations of vectors $e'_{\beta'}$ are dense in $\cal L'.$ (In other words these systems of vectors are overcomplete.)

Let us fix a non-degenerate pairing $\bra l,l'\ket  $ between $\cal L$ and $\cal L'.$ (We can consider either bilinear pairing or a pairing that is linear with respect to one argument and antilinear with respect to the second argument.)

We assume that
\begin{equation}\label{O}
\bra l,l'\ket=\int_ {\cal M\times\cal M'} dm dm'\bra l,e'_{ m'}\ket R(m,m')\bra e_m,l' \ket,
\end {equation}
where $dm dm'$ is the measure on $\cal M\times\cal M'$. In different notations, this formula can be written as 
\begin {equation} \label {RR} 
\int dm dm' \ |e'_{m'}\ket R(m,m')\bra e_m|=1.
\end {equation}

We define covariant symbol  $\uu A(\alpha, \beta')$  of operator $A$ acting in $\cal L$ by the formula
\begin{equation}\label {CO}
\uu A(\alpha, \beta')=\frac{ \bra Ae_{\alpha},e'_{\beta'}\ket} { \bra e_{\alpha},e'_{\beta'}\ket}.
\end{equation}
In bra-ket notations

\begin{equation}\label {COO}
\uu A(\alpha, \beta')=\frac{ \bra e'_{\beta'}|A|e_{\alpha}\ket} { \bra e'_{\beta'}| e_{\alpha}\ket}.
\end{equation}

In particular, we can assume that $\cal L$ $=\cal L'$ is a Fock space (Hilbert space of Fock representation of canonical commutation relations) and the overcomplete system of vectors in this space coinsists of eigenvectors of annihilation operators (of Poisson vectors).  Then the covariant symbol coincides with Wick symbol. 

 Recall that the Wick symbol can be defined in the following way.  Represent the operator in normal form (creation operators $\hat a^*(f)$ from the left, annihilation operators $\hat a(g)$ from the right). Remove hats.
 Resulting polynomial of $a^*, a$ is a Wick symbol of the operator.

Notice that the spaces $\cal L$ and $\cal L'$ are on equal footing in our construction; hence we can define the covariant symbol of an operator $B$ acting in $\cal  L'$ in a similar way. We say that operators $A$ and $B$ are dual if $\bra Ax,y\ket =\bra x, By \ket,$ it is easy to check that symbols of dual operators  coincide

\begin{equation}\label {DU}
\uu B(\alpha, \beta')= \frac{ \bra e_{\alpha}, B e'_{\beta'}\ket} { \bra e_{\alpha},e'_{\beta'}\ket}={\uu A}(\alpha,\beta').
\end{equation}

The covariant symbol $\uu C=\uu A*\uu B$ of operator $C=AB$ is given by the formula
\begin{equation}\label {PR}
\uu C(\alpha, \beta')=
\int dmdm' \ \uu B(\alpha,m')\uu A(m,\beta')R(m,m')\frac {\bra e_{ \alpha}, e'_{m' }\ket \bra e_ m,e'_{\beta'}\ket}{\bra e_{\alpha},e'_{\beta'}\ket}.
\end{equation}

This formula agrees with (\ref {NN}) if we take  
$$\bra e_{m}, e'_{m' }\ket =e^{c(m,m')}.$$

We define  contravariant symbol 
$\mathring{A}(\alpha, \beta')$ 
  of operator $A$ acting in $\cal L$ by the formula

\begin{equation}\label {CR}
A e_{\alpha}=\int d\gamma d\gamma' \mathring{A}(\gamma,\gamma')\bra e_{\alpha},e'_{\gamma'}\ket R(\gamma,\gamma')e_{\gamma}.
\end{equation}

It is easy to express covariant symbols in terms of contravariant symbols
$$\uu A(\alpha,\beta')=\int d\gamma d\gamma' \mathring{A}(\gamma,\gamma')
\frac {\bra e_{\alpha},e'_{\gamma'} \ket R(\gamma,\gamma')\bra e_{\gamma},e'_{\beta'}\ket}
{\bra e_{\alpha},e'_{\beta'}\ket}.$$
To calculate the contravariant symbol $\mathring{C}$
 of the product $C=AB$  of operators $A,B$ we notice that
\begin{equation}
C e_{\alpha}=\int d\gamma_1 d\gamma_1' d\gamma_2 d\gamma_2' \mathring{A}(\gamma_2,\gamma_2')\bra e_{\alpha},e'_{\gamma_2'}\ket 
R(\gamma_2,\gamma_2') \mathring{B}(\gamma_1,\gamma_1')\bra e_{\gamma_2},e'_{\gamma_1'}\ket R(\gamma_1,\gamma_1')e_{\gamma_1}.
\end{equation}
Hence  $\mathring{C}$ can be expressed in terms of  contravariant symbols of factors by the formula 

\begin{equation}
\mathring{C}(\gamma,\gamma')= \frac{1}{R(\gamma,\gamma')}  \int  d\beta d\beta'  \bra e_{\beta},e'_{\beta'}\ket \mathring{A}(\beta,\gamma') R(\beta,\gamma') \mathring{B}(\gamma,\beta')R(\gamma,\beta').
\end{equation}

 This expression has the form (\ref {NN1}) with
 $$e^{c(\gamma,\gamma')}=R(\gamma,\gamma'),\;\;\;e^{-r(\gamma,\gamma')}= \bra e_{\gamma},e'_{\gamma'}\ket.
 $$

 Notice that a bounded operator always has a covariant symbol. Moreover, an operator $A$ has a covariant symbol if it is unbounded but the vectors $e_{\alpha}$ belong to the domain where $A$ is defined. However, it is non-trivial to say whether there exists an operator with a given covariant symbol. For contravariant symbols, the situation is the opposite.  It is easy to give conditions for the existence of an operator with a given contravariant symbol, but if we know an operator it is non-trivial to say whether is has a contravariant symbol.
 
\section { Coherent states}

As in Section 3 we consider   Banach spaces $\cal L$ and $\cal L'$, non-degenerate pairing $\bra l, l' \ket$ between these spaces ,  and systems of vectors
$e_{\alpha}\in \cal L$, $e'_{\beta'}\in \cal L'.$ Here  $\alpha\in \cal M, \beta'\in\cal M'$.

We fix a representation $T$ of Lie group $G$ in the space $\cal L$ and   representation $T'$ in the space   $\cal L'$  in such a way that  $G$ acts transitively on the set of vectors $e_{\alpha}$
and on the set of vectors  $e'_{\beta'}$.  This means that $\cal M$ and $\cal M'$  can be considered as homogeneous spaces: $\cal M$ $=G/H$ and $\cal M'$ $=G/H'.$ We assume that the representations $T'$ and $T$ are dual: $\bra Tl,l'\ket =\bra l,T'l'\ket.$

We require that linear combinations of vectors $e_{\alpha}$ are dense in $\cal L$ and linear combinations of vectors $e'_{\beta'}$ are dense in $\cal L'$; it follows that representations $T$ and $T'$ are irreducible.

To define covariant and contravariant symbols starting with vectors $e_{\alpha}, e'_{\beta'}$ 
we need the relation (\ref{NN}). If there exist a $G$- invariant measure  $dmdm'$ on  $\cal M$  $\times \cal M'$ one can obtain such a relation taking $R=1$.  To prove this fact we notice that  the expression
\begin {equation} \label {RRR} 
   \int dm dm' \ |e'_{m'}\ket \bra e_m|
\end {equation}   
 is $G$-invariant.  The integrand of (\ref {RRR})  specifies an operator in $\cal L$; we assume that  the integral is converging, hence (\ref {RRR}) can be regarded as an operator in $\cal L$ commuting with all operators $T(g), g\in G. $ It follows from the irreducibility of the representation $T$ that (\ref {RRR}) is a constant.  Multiplying the measure $dmdm'$ by a constant factor we obtain (\ref {RR}).
 
 If $\cal M$ $=\cal M'$ and there exists a $G$-invariant measure $dm$ on $\cal M$ then 
 we get the relation (\ref {NN}) with $R=\delta (\alpha,\beta')$ (assuming  convergence of  the integral). Using formulas (\ref {SS4}) and (\ref {S444}) we obtain functional integrals in this situation.

 One says that the vectors $e_{\alpha}$ and $e'_{\beta'}$  considered in the present section are coherent states. This definition 
generalizes  the definition of coherent state in $\cite {PER}$ where $\cal L$ $=\cal L'$ is a Hilbert
space, $T$ and $T'$ are unitary operators, $e_{\alpha}=e'_{\alpha}$. Notice, that in (\cite {PER})
the group $G$ transforms the vector $e_{\alpha}$ in a vector, proportional to the vector of the same kind (it acts transitively on corresponding elements of projectivization of $\cal L$).  In our setting, we also can consider a similar situation.

\section{L-functionals}
\subsection{ First definition}

Let us consider a unital associative  algebra   with generators $\gamma (f)$ obeying canonical commutation relations  (CCR):
\begin {equation} \label {CCR}
\gamma (f)\gamma(g)-\gamma (g)\gamma (f)= i (f,g)
\end{equation}
 (Weyl algebra). 

Here $f$ and $g$  are elements of real vector space $E$ equipped with non-degenerate antisymmetric inner product $(\cdot,\cdot)$, generators $\gamma(f)$ depend linearly of $f$. We assume that Weyl algebra is a complex algebra equipped with an antilinear involution  and
generators $\gamma(f)$ are self-adjoint with respect to this involution.

Let us fix a representation of the Weyl algebra (representation of canonical commutation relations) in Hilbert space $\cal F$. We assume that generators are represented by self-adjoint operators $\hat \gamma(f)$; hence we can consider unitary operators $V_f=\exp(i\hat \gamma(f))$. It is easy to check that
\begin {equation} \label {WE}
V_fV_g=V_{f+g}\exp(\frac{i}{2}(f,g)).
\end {equation}

These relations are formally equivalent to (\ref {CCR}). We consider the smallest linear subspace of the space of bounded linear operators in $\cal F$ containing all operators $V_f$; the closure of this space in norm-topology is a $C^*$-algebra that can be regarded as an exponential form of Weyl algebra (see, for example,\cite {H} and references therein for the mathematical theory of Weyl algebra).
We will work with this algebra denoted  by $\cal W.$ The space of continuous linear functionals on $\cal W$ will be denoted by $\cal L.$ Notice that a functional $L\in \cal L$ is determined by its values  on   operators $V_f$, therefore we 
can consider $L$ as a non-linear functional  ${\bf L}(f)=L(V_f)$ on $E$ (the representation of states of Weyl algebra by means of non-linear functionals was rediscovered and studied in \cite{H}).

In particular, positive functionals on the algebra $\cal W$ (quantum states) can be represented by non-linear functionals; we will use the term L-functional for non-linear functionals representing states.  (Recall that a linear functional $L$ on $*$-algebra $\cal A$ is positive if 
$L({A}^*A)\geq 0$ for every $A\in \cal A.$)  

If we have a normalized vector $\Phi$ or, more generally, a  density matrix $K$ in representation space of  some $*$-algebra $\cal A$ we can obtain a quantum state 
$\omega$ by the formulas

\begin {equation}\label {LP}
\omega (A)=\bra \Phi, \hat A\Phi\ket,
\end{equation}
\begin {equation}\label {LK}
\omega (A)=Tr \hat AK,
\end{equation}
where $\hat A$ stands for the operator representing an element $A\in \cal A.$

Every quantum state can be represented by a vector in some representation of $*$-algebra (Gelfand-Naimark-Segal construction).

If $\cal A$ is the Weyl algebra $\cal W$ we represent a density matrix $K$ in any representation of canonical commutation relations (= in any representation of $\cal W$) by L-functional

 $${\bf L}_K(f)=TrV_fK.$$
 One can say that  L-functionals describe states in all representations of canonical commutation relations.
 
The evolution operators of quantum theory constitute a one-parameter group of automorphisms of the algebra $\cal W$ generated by an infinitesimal automorphism $H.$  They induce evolution operators acting on quantum states; these operators can be extended to $\cal L$. To find evolution operators one should solve the equation of motion (\ref {EM}). We apply the methods of preceding sections assuming that $\cal L'$ \ $=\cal W.$ We define covariant symbols of operators acting in $\cal L$
using systems of vectors $e_f\in \cal L$  and vectors $e'_{f'}\in \cal L'$ that are defined in the following way. We assume that $f,f'\in E$, $e_f(V_g)=\exp{\frac{i}{2}(f, g)}, e'_{f'}=V_{f'}$. It follows that
$\bra e_f, e'_{f'}\ket={\exp{\frac{i}{2}(f, f')}}.$ To  get a function $R$ obeying (\ref {O}) we  can  take  
${R(f, f')= C\exp{(-\frac{i}{2}(f, f'))} }$ where the constant $C$  is chosen in such a way that

\begin{equation} \label {unty}
\int{dfdf'} \bra e_g, e'_{f'}\ket R(f, f')\bra e_f, e'_{g'}\ket=\bra e_g, e'_{g'}\ket.
\end{equation}

Here $dfdf'$ is a measure on $E\times E$ or at least a rule that allows us to calculate integrals of some functions defined on this space (in (\ref{unty}) we need only integrals of quadratic exponents).

In what follows we assume  that the   antisymmetric inner product is represented in the form $(f,g)=f^i\sigma_{ij}g^j=f\sigma g$ and $\dim E=2n<\infty$; then  
{$C={|det(\sigma)|^{1/2}}/{(2\pi)^n}.$}
We assume that $C=1$  by changing the measure $dfdf'.$ Then  in the notations of Section 2 we have $c(f,f')=r(f,f')= {\frac{i}{2}(f, f')}.$

Let us suppose that the evolution is specified by an infinitesimal automorphism of Weyl algebra $\cal W$  $=\cal L'$ represented as a
commutator $H$  of the element of $\cal W$  with $i \hat H$.  Here $\hat H$ is a self-adjoint element of $\cal W$:

\begin{equation} 
\hat H=\int d\beta h(\beta) V_{\beta},
\end{equation}
where  $h(-\beta)=\bar{h}(\beta).$ (Notice that  one can  consider also a more general case when $\hat H$ is a formal expression such that the commutator with $\hat H$ makes sense.)

It is easy to check that the covariant symbol of the operator $H$ has the form
\begin{equation} \label {uuH}
\uu H(f,f')= i\int d\beta h(\beta) (e^{\frac{i}{2}((f,\beta)-(\beta,f'))}-e^{\frac{i}{2}((f,\beta)+(\beta,f'))})=
2\int d\beta h(\beta) e^{\frac{i}{2}(f,\beta)} \sin {\frac{\scriptstyle{(\beta,f')}}{{2}}}.
\end{equation}

Using (\ref{I}) we obtain  a representation  of the symbol of the  evolution operator  in ${\cal W}=\cal L'$  in terms of functional integrals
\begin {equation} \label {SSS3}
\begin{gathered}
\uu {e^{tH}}(f,f')= \int \prod df(\tau) df'(\tau)e^{S[f(\tau), f'(\tau)]},\\
 S[f(\tau), f'(\tau)]=\int_0^t \left(\uu H(f(\tau), f'(\tau))-\frac{i}{2}(f(\tau), \dot f'(\tau)))\right)d\tau +\frac{i}{2}((f(t),f')-(f,f')),
\end{gathered}
  \end {equation}
where we integrate over the set of functions obeying conditions $f(0)=f; f'(t)=f'$. 

The evolution operator $e^{tH}$ in the algebra $\cal W$ is dual to the evolution operator $e^{tK}$ in the space $\cal L$ of linear functionals on $\cal W:$
$$\bra e^{tK} x, y\ket=\bra x, e^{tH}y\ket,$$
hence the operator $K$ entering the equation of motion for $L$-functionals is dual  to the infinitesimal  automorphism $H$. Using the formula (\ref {DU}) we can say the symbol of $K$ coincides with the symbol of $H$ and the symbols of operators of evolution $e^{tK}$ and $e^{tH}$ coincide. This remark allows us to say that {\it  the symbol of the operator of evolution in the formalism of L-functionals is expressed in terms of functional integrals by the formula} (\ref {SSS3}).

The same statement can be obtained from the equation of motion in the formalism of L-functionals.
The time derivative of ${\bf L}(f)=L(V_f)$  can be written in the form

$$
i\frac {dL(V_f)}{dt}=  \int d\beta h(\beta) L(V_fV_{\beta}-V_{\beta}V_f)=\int d\beta h(\beta)(e^{\frac{i}{2}f\sigma\beta}-e^{\frac{i}{2}\beta\sigma f})L(V_{f+\beta}),$$
hence
\begin{equation} \label {LEM}
\frac {d{\bf L} (f)}{dt}=2\int d\beta h(\beta){\sin(\frac{1}{2}f\sigma\beta)}{\bf L}(f+\beta).
\end{equation}

We can write (\ref {LEM}) in the form 
$$\frac {d{\bf L} (f)}{dt}= K{\bf L}(f),$$
where
$$(K{\bf L})(f)=2\int d\beta h(\beta) {\sin(\frac{1}{2}f\sigma\beta)}(T_{\beta}{\bf L})(f).$$

Here $(T_{\beta}{\bf  L})(f)={\bf L}(f+\beta).$

It is easy to check that the symbol of the operator $K$ is given by the formula (\ref {uuH}). We obtain another derivation of the functional integral  (\ref {SSS3}) for the evolution operator in the formalism of $L$-functionals.
\vskip .1in
Sometimes it is convenient to introduce the Planck constant $\hbar$ in 
the formula (\ref{LEM}) assuming that in the defining relations of Weyl algebra we have $\hbar$ in the right-hand side:
\begin {equation}\nonumber
\gamma (f)\gamma(g)-\gamma (g)\gamma (f)= i\hbar(f,g),
\end{equation}
where $(f,g)=f\sigma g=f^i\sigma_{ij}g^j$
and replacing $\hat H$ with $ {\hat H}/{\hbar}.$
Then 
\begin{equation} \label {LEMM}
\frac {d{\bf L} (f)}{dt}=\int d\beta h(\beta)\frac {2\sin(\frac{\hbar}{2}f\sigma\beta)}{\hbar} {\bf L}(f+\beta).
\end{equation}

It follows from (\ref {LEMM}) that the equation of motion for $L$-functionals has a limit as $\hbar \to 0.$

\subsection{Second definition}

Let us consider another form of  canonical  commutation relations: 
\begin {equation}\label {CCC}
[a(f),a(f')]= [a^*(g),a^*(g')]=0,\;\;\;\; [a(f),a^*(g)]= (f,g),
\end{equation}
where $(f,g)$ is a non-degenerate pairing between vector space $\cal E$ and complex conjugate vector space $\bar {\cal E}$. (Here  $f,f'\in{ \cal E}, g,g'\in\bar{ \cal E}$.) We will assume that this paring is defined on $E$; then it is linear with respect to the first argument and antilinear with respect to the second argument.(Notice that in our notations $E$ and $\bar E$ consist of the same elements but have different complex structures.) Then (\ref {CCC}) takes the form
\begin {equation}\label {CCCC}
[a(f),a(f')]= [a^*(g),a^*(g')]=0,\;\;\;\; [a(f),(a(g))^*]=(f,g).
\end{equation}
(The involution $^*$ transforms $a(f)$ into $a^*(f^*).$ 
We assume that $a(f)$
is linear with respect to $f$, then  $a^*(f^*)= (a(f))^*$ is antilinear with respect to $f\in E$.)  We do not assume that
the pairing $(f,g)$ is well-defined for all pairs $f,g$; in particular,
 $(f,f)$ can be infinite.)

If  the space $\cal E$ consists of functions on measure space $\cal M$ then $a(f),a^*(g)$ should be regarded as 
generalized functions: $a(f)=\int f(k)a(k)dk,\;\; a^*(g)=\int {g}(k)a^*(k)dk.$
Then  canonical commutation relations (\ref{CCC}) can be written in the form

\begin {equation}\label {CCCk}
[a(k),a(k')]= [a^*(k),a^*(k')]=0,\;\;\;\; [a(k),a^*(k')]=\delta(k,k').
\end{equation} 
If $k$ is a discrete parameter (i.e.$\cal M$ is a discrete set with counting measure) the above relations can be written as follows
$$ [a_k,a_{k'}]=[a^*_k,a^*_{k'}]=0,\;\;\;\; [a_k,a^*_{k'}]=\delta_{k,k'}.$$

The relations (\ref{CCC}) are obviously equivalent to the relations (\ref {CCR}) (to get (\ref {CCR}) from (\ref{CCC}) we can consider self-adjoint elements $a(f)+a^*(f), i(a(f)-a^*(f)$).

The relations (\ref{CCC}) are especially convenient in the case of an infinite number of degrees of freedom. In this situation one should use the original definition of L-functional (see \cite {SCH}, \cite {S}, \cite {MO}.)

Again we can write canonical commutation relations in exponential form introducing expressions 
$$W_{\alpha}=e^{-a^*(\alpha)}e^{a(\alpha^*)}.$$
Notice that $W_{\alpha}$ is not holomorphic with respect to $\alpha$ therefore it would be more appropriate to use the notation $W_{\alpha^*,\alpha}$ as we are doing in similar situations below.

It is easy to check that
\begin {equation}\label {EX}
W_{\alpha}W_{\beta}= e^{-(\alpha^*,\beta)} W_{\alpha+\beta} .
\end{equation}

Notice that in the case when $(\alpha,\alpha)$ is finite $W_{\alpha}$ coincides with $V_{f}$ up to a finite constant factor.   However, we do not assume that $(\alpha,\alpha)<\infty.$ (This is important for applications to string theory.)

We define vector space $\cal W$ as a space of linear combinations of expressions of the form $P_{\alpha}W_{\alpha}$ where $\alpha\in E$ and $P_{\alpha}$ belongs to some class of polynomials with respect to $a^*,a.$
The relations (\ref{CCC}), (\ref {EX})  specify multiplication in $\cal W$, but this multiplication is not always defined. Nevertheless one can consider $\cal W$ as a version of Weyl algebra. (Better to say that our construction gives various versions of Weyl algebra because we did not specify the class  of polynomials and topology in $\cal W.$) We fix some topology in $\cal W$ in such a way that $W_{\alpha}$ is infinitely differentiable with respect to $\alpha, \alpha^*.$
Then the elements $W_{\alpha}$ are dense in $\cal W$ (diffrentiating $W_{\alpha}$ we obtain polynomials of $a^*,a$). This means that a continuous linear functional $L$ on $\cal W$ is specified by non-linear functional ${\bf L}(\alpha^*,\alpha)=L(W_{\alpha})$ (by values of $L$ on $W_{\alpha}).$ The space of of continuous linear functionals on $\cal W$  is denoted by $\cal L.$ Notice that we can say that
${\bf L}(\alpha^*,\alpha)=\bra L,W_{\alpha}\ket$ where $\bra \cdot,\cdot \ket$ stands for the standard pairing between $\cal L$ and $\cal W.$

We say that non-linear functionals corresponding to quantum states (to positive functionals =elements of $\cal L$ obeying $L({A}^*A)\geq 0$) are L-functionals.  

We represent a density matrix $K$ in any representation of canonical commutation relations  (\ref {CCC}), (\ref {EX}) by L-functional

 $${\bf L}_K(\alpha^*, \alpha)=TrW_{\alpha}K= Tre^{-a^*(\alpha)} e^{a(\alpha^*)}K.$$

Every element $B\in \cal W$  specifies two operators acting in the space of linear functionals $\cal L.$ The first operator transforms the functional $\omega (A)$ into the functional $\omega(AB).$ Applying this construction to the cases $B=a(f)$ and $B=a^*(f)$  we obtain operators denoted by  $b^+(f)$ and $b(f)$.
The second operator transforms the functional  $\omega(A)$ into the functional $\omega(B^*A)$; if we start with $B=a(f), B=a^*(f)$ we get operators denoted by $\tilde b(f), \tilde b^+(f)$.

The evolution operators of quantum theory constitute a one-parameter group of automorphisms of  $\cal W$ generated by an infinitesimal automorphism $H.$  They induce evolution operators acting on $\cal L$ and transforming quantum states to quantum states. 

Let us suppose that the evolution is specified by an infinitesimal automorphism of Weyl algebra $\cal W$  $=\cal L'$ represented as a
commutator $H$  of the element of $\cal W$  with $\frac {1}{i} \hat H$.  Here $\hat H$ is a self-adjoint element of $\cal W$
or a self-adjoint formal expression
\begin{equation}\label {HH}
\hat H=\sum_{m,n}\int  H_{m,n}(k_1,...k_m|l_1,...,l_n)a^*(k_1)...a^*(k_m)a(l_1)...a(l_n)\prod_{
\genfrac {}{}{0pt}{2}{1\leq i\leq m}{1\leq j\leq n}
} dk_i dl_j 
\end {equation}
such that the commutator $H$ is a well-defined derivation of $\cal W$ that can be regarded as an infinitesimal automorphism (i.e. solving the equations of motion we obtain a one-parameter group of evolution operators $e^{tH}$). This allows us to write an equation  of motion (\ref{EM})  in the space $\cal L$ taking $H=H_L-H_R$ where 

$$ H_R=
\sum_{m,n}\int  H_{m,n}(k_1,...k_m|l_1,...,l_n)b^+(k_1)...b^+(k_m)b(l_1)...b(l_n)\prod_{
\genfrac {}{}{0pt}{2}{1\leq i\leq m}{1\leq j\leq n}} dk_i dl_j, 
$$
$$ H_L=
\sum_{m,n}\int  H_{m,n}(k_1,...k_m|l_1,...,l_n)\tilde b^+(k_1)...\tilde b^+(k_m)\tilde b(l_1)...\tilde b(l_n)\prod_{
\genfrac {}{}{0pt}{2}{1\leq i\leq m}{1\leq j\leq n}} dk_i dl_j .
$$

We solve the equation of motion (\ref {EM}) applying  the methods of  Section 2 and assuming that $\cal L'$ \ $=\cal W.$ We define covariant symbols of operators acting in $\cal L$
using systems of vectors $e_f\in \cal L$  and vectors $e'_{f'}\in \cal L'$ that are defined in the following way. We assume that  $e_f\in \cal L$ corresponds to a non-linear functional $e_f(W_{\alpha})=\exp i((f,\alpha^*)+(\alpha,f^*))$ and that $e'_{f'}=W_{f'}$. It follows that
$\bra e_f, e'_{f'}\ket=\exp i((f,f^{*'})+(f',f^*)).$ To  get a function $R$ obeying (\ref {O}) we    take  
$R(f, f')= C\exp (-i(f,f^{*'})-i(f',f^*))$ where the constant $C$  is chosen in such a way that
the formula(\ref{unty}) is satisfied.

It is easy to calculate the covariant symbol of the  operator $H:$
$$
\begin{gathered}
 \uu H(f,f') =
i\sum_{m,n}\int\prod_{
\genfrac {}{}{0pt}{2}{1\leq i\leq m}{1\leq j\leq n}} (dk_i dl_j) 
 H_{m,n}(k_1,...k_m|l_1,...,l_n)\times
\\
 \left(\; f^*(k_1)...f^*(k_m)f'(l_1)...f'(l_n)-f'^*(k_1)...f'^*(k_m)f(l_1)...f(l_n)\;\right).
\end{gathered}
$$

This allows us to get a representation of the symbol of the evolution operator in terms of functional integrals
\begin {equation} \label {SSS5}
\begin{gathered}
\uu {e^{tH}}(f,f')= \int \prod df(\tau) df'(\tau)e^{S[f(\tau), f'(\tau)]},\\
 S[f(\tau), f'(\tau)]=\int_0^t \left(\uu H(f(\tau), f'(\tau))
 -i(f(\tau), \dot f'^*(\tau))-i(f^*(\tau), \dot f'(\tau))\right)d\tau + 
\\ +i((f(t),f'^*)+(f^*(t),f')-(f,f'^*)-(f^*,f')).
\end{gathered}
\end {equation}

In particular, if $\hat H=\int dk\epsilon(k)a^*(k)a(k)$ is a quadratic translation-invariant  Hamiltonian we obtain
$$ \uu H(f,f') =
i\int dk \epsilon(k) \left(f^*(k)f'(k)-f'^*(k)f(k)\right).
$$

Here $k\in \mathbb {R}^d.$  We  can consider also a more general case when $k=({\bf  k},s)$ where $s$ is a discrete index and
 ${\bf k}\in \mathbb {R}^d.$
 
 Let us consider a general translation-invariant Hamiltonian (\ref {HH}). In other words, we assume that the integrand in (\ref{HH})  contains delta-function $\delta(\bf {k_1+...k_m-l_1-...-l_n})$  (the arguments $l,l$ are points of $\mathbb{R}^d$ plus discrete indices).
 
 We represent $\hat H$ as a sum of the quadratic part $\hat H_0$ and perturbation $g\hat V.$
 Then we can consider time-dependent Hamiltonian $\hat H(t)=\hat H_0+h(at)g\hat V$ where 
 $h(0)=1, h(-\infty)=0.$ For $a\to 0$ this means that we switch on the interaction $\hat V$ adiabatically. If $U_a(t, -\infty)$ denotes the evolution operator for the corresponding "Hamiltonian" $H(t)$ and $\Phi$ stands for a translation-invariant stationary state of quadratic "Hamiltonian" $H_0$. Then $\Psi=\lim_{a\to 0}U_a(0,\infty)\Phi$ is a translation-invariant stationary state of the "Hamiltonian" $H_0+gV.$ 
 
In the derivation of the  formula (\ref {SSS5}) we assumed that the "Hamiltonian" $H$ does not depend on time but this formula  can be applied also to time-dependent "Hamiltonians". This remark allows us to express $\Psi$ in terms of functional integrals.  If we start with an equilibrium state $\Phi$  the state $\Psi$ is also an equilibrium state (in general with different temperature), however, the above considerations can be applied also in non-equilibrium situations. They can be considered as justification of Keldysh formalism in non-equilibrium statistical physics and lead to the same Feynman diagrams. (See \cite{S} for another derivation of Keldysh diagram  techniques in the formalism of L-functionals.)

The above formulas were written in the assumption that $\hbar=1.$  In general we should include the factor $\hbar$ into the right-hand side of the formula (\ref {CCC}) and into the left-hand side of the equation of motion (\ref {EM}).

 Itf we represent elements of $\cal L$ by non-linear functionals $\bf L$
 the operators $b(f), b^+(f)$,  $\tilde b(f), \tilde b^+(f)$  can be represented in the form

$$b^+(k)=-\hbar \alpha_k+\frac{\partial}{\partial \alpha^*_k} \equiv -\hbar c_2^*(k)+c_1(k), \;\;\; b(k)=-\frac{\partial}{\partial \alpha_k} \equiv -c_2(k), $$
$$\tilde b^+(k)=\hbar \alpha^*_k-\frac{\partial}{\partial \alpha_k} \equiv \hbar c^*_1(k)-c_2(k),
\;\;\;\tilde b(k)=\frac{\partial}{\partial \alpha^*_k}\equiv c_1(k),$$
where $ c^*_i (k)$ are operators of multiplication by $\alpha_k^*$ for $i=1$ and by $\alpha_k$ for $i=2$, and $c_i(k)$ are derivatives taken, respectively, with respect to $\alpha_k^*$ and $\alpha_k$.  (To simplify  notations we  assumed that  $E$ consists of functions on discrete space $\cal M$; points of $\cal M$ are labeled by index $k.$)

It is easy to derive from these formulas that the equations of motion for functionals  ${\bf L}(\alpha^*,\alpha)$ have a limit as $\hbar$ tends to zero.

\subsection {Clifford algebra}

Clifford algebra is defined by canonical anticommutation relations

\begin {equation}\label {CCA}
[a(f),a(f')]_+= [a^*(g),a^*(g')]_+=0,\;\;\;\; [a(f),a^*(g)]_+= (f,g)
\end{equation}

 In other words to define Clifford algebra we take the definition of Weyl algebra and replace commutators with anticommutators.
 
 The results above can be generalized to Clifford algebra. The main difference is that the symbols should be considered as functions of anticommuting variables.

 The simplest way to understand this is to notice that Clifford algebra can be regarded as super Weyl algebra.
 
 Recall that for ${\mathbb{Z}}_2$-graded space $E$ and Grassmann algebra $\Lambda$ one can define a $\Lambda$-point of $E$ as a formal linear combination $\sum\lambda_Ae_A$ where $e_A$ is a basis of $E$ and the coefficients $\lambda_A$ are even for even $e_A$, odd for odd $e_A$. The set $E(\Lambda)$ of $\Lambda$- points can be regarded as vector space. 
 If $E$ is an algebra this set also can be considered as an algebra. If for all Grassmann algebras $\Lambda$ the set $E(\Lambda)$ is a Lie algebra one says that $E$ is a super Lie algebra. Similarly if for all Grassmann algebras the set $E(\Lambda)$ is a Weyl algebra one can say that $E$ is a super Weyl algebra. In particular,  if f ${\mathbb{Z}}_2$-graded space $E$ is purely odd and equipped with a structure of  Clifford algebra then the set of $\Lambda$-points is a Weyl algebra. This means that Clifford algebra can be considered as  super Weyl algebra.

{\bf Acknowledgements} The authors are indebted to A. Dynin, B. Mityagin,  A. Neklyudov, A. Rosly,
V. Zagrebnov for useful comments.

\end{document}